\documentclass{article}

\usepackage{lineno}

\usepackage{spconf,amsmath,graphicx}
\usepackage{caption}               
\usepackage{subcaption}
\usepackage{times}
\usepackage{float}
\usepackage{cuted}                 
\usepackage{amssymb}
\usepackage{booktabs}
\usepackage{multirow}
\usepackage{algpseudocode,algorithm}
\usepackage{comment}
\usepackage{pifont}
\usepackage{enumitem}
\usepackage{diagbox}
\usepackage[table]{xcolor}
\definecolor{bestgreen}{HTML}{C6EFCE}    

\definecolor{cvprblue}{rgb}{0.21,0.49,0.74}
\definecolor{BLUE}{rgb}{0.015,0.345,0.576}
\definecolor{ORANGE}{rgb}{0.858,0.380,0.0}
\definecolor{GREEN}{rgb}{0.062,0.502,0.062}

\usepackage[pagebackref,breaklinks,colorlinks,citecolor=cvprblue]{hyperref}

\newcommand{\cmark}{{\color[HTML]{34FF34}\textbf{\ding{51}}}}%
\newcommand{\xmark}{{\color[HTML]{FE0000}\textbf{\ding{55}}}}%
\title{\textcolor{BLUE}{Vol}\textcolor{ORANGE}{Hu}\textcolor{GREEN}{Me}: a High-Resolution Large Scale Dataset of \\ \textcolor{BLUE}{Vol}umetric \textcolor{ORANGE}{Hu}man \textcolor{GREEN}{Me}shes}
\name{%
  Giulia Martinelli$^{1,2}$,%
  \quad Niccolò Bisagno$^{1,2}$,%
  \quad Nicola Garau$^{1,2}$,%
  \quad Esa Rahtu$^{3}$,%
  \quad Nicola Conci$^{1,2}$
}
\address{%
  $^{1}$University of Trento, Trento, Italy,
  $^{2}$CNIT, Italy,
  $^{3}$Tampere University, Tampere, Finland 
\vspace*{-50pt}
}

\begin{document}
  
\maketitle

\begin{strip}
\centering
    \includegraphics[width=0.8\textwidth]{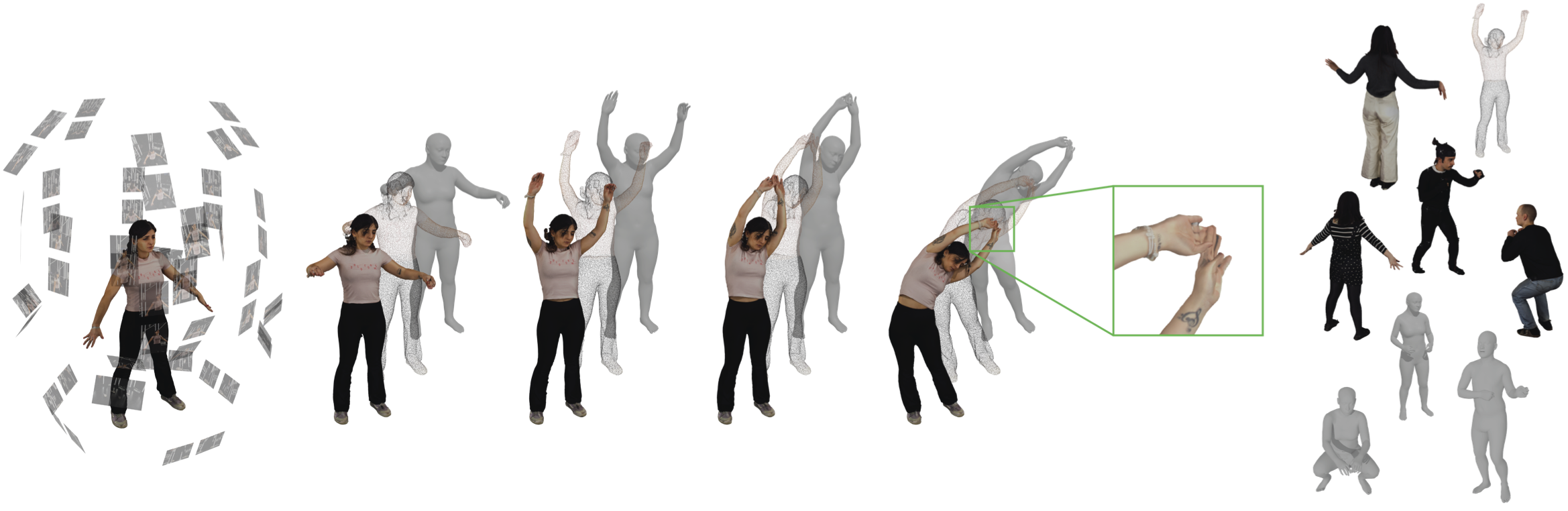}
    \captionof{figure}{Existing volumetric human datasets do not preserve close-range, high-resolution body part details, usually relying on distant, full-body captures, with only a subset of annotations. Our dataset, VolHuMe, is designed to preserve close-range, high-resolution body-part details. We provide a complete set of annotations, such as multi-view RGB–D, high-resolution meshes, dense point clouds, and SMPL-X fittings. Our data enables a variety of tasks, namely 4D reconstruction, point-cloud estimation, and novel-view rendering.}
    \label{fig:teaser}
\end{strip}

\begin{abstract} 
\vspace*{-5pt}
We introduce VolHuMe, a dataset of high-quality 4D human scans captured with a state-of-the-art volumetric studio using 64 RGB and 32 depth cameras. VolHuMe contains individual captures of 104 subjects and provides extensive ground truth, including SMPL-X, high-resolution meshes, multi-view RGB/depth images, rigged meshes, point clouds, garment segmentation, and detailed hand and facial geometry.
Unlike prior datasets that primarily rely on full-body imagery, VolHuMe uses a close-range, high-resolution capture setup that preserves fine-grained body-part details, improving geometric fidelity and texture resolution. We benchmark VolHuMe on state-of-the-art methods across 3D and 4D human reconstruction tasks, showcasing the dataset’s quality and exposing the limitations of current evaluation testbeds.
\end{abstract}
%
\section{Introduction}
\label{sec:intro}

\begin{table*}[]
\resizebox{0.98\textwidth}{!}{%
\begin{tabular}{ccccccccc}
\hline
 & \textbf{Dataset} & \textbf{\begin{tabular}[c]{@{}c@{}}\# of\\ unique\\ subjects\end{tabular}} & \begin{tabular}[c]{@{}c@{}}Full body\\ point cloud\end{tabular} & \begin{tabular}[c]{@{}c@{}}High-res\\ mesh\end{tabular} & Textured & \begin{tabular}[c]{@{}c@{}}Normal\\ maps\end{tabular} & Rigged & Fitting \\ \hline
\multirow{3}{*}{\rotatebox{90}{\begin{tabular}[c]{@{}c@{}}In-the\\ wild\end{tabular}}} & Agora \cite{Patel2021agora} & 350  & \xmark & \xmark & \cmark & \xmark & \xmark & SMPL-X \\
 & BEDLAM \cite{black2023bedlam} & 271 & \xmark & \xmark & \cmark & \xmark & \xmark & SMPL-X \\
 & EMDB \cite{kaufmann2023emdb} & 10 & \xmark & \xmark & \xmark & \cmark & \xmark & SMPL \\ \hline
\multirow{10}{*}{\rotatebox{90}{Controlled}} & Human3.6M \cite{ionescu2013human3} & 11 & \cmark & \xmark & \xmark & \xmark & \xmark & \xmark \\
 & CMUPanoptic \cite{joo2015panoptic} & ~40 & \cmark & \xmark & \xmark & \xmark & \xmark & \xmark \\
 & Human4D \cite{chatzitofis2020human4d} & 4 & \cmark & \xmark & \cmark & \cmark & \cmark & \xmark \\
 & HUMBI \cite{yoon2021humbi} & 772 & \cmark & \cmark & \cmark & \xmark & \xmark & SMPL \\
 & THuman4.0 \cite{zheng2022structured} & 3 & \xmark & \xmark & \cmark & \xmark & \xmark & SMPL-X \\
 & AvatarRex \cite{zheng2023avatarrex} & 4 & \xmark & \xmark & \xmark & \xmark & \xmark & SMPL-X \\
 & ActorsHQ \cite{icsik2023humanrf} & 8 & \xmark & \cmark & \cmark & \xmark & \xmark & \xmark \\
 & DNA-Rendering \cite{cheng2023dna} & 500 & \cmark & \xmark & \cmark & \xmark & \xmark & SMPL-X \\
 & MVHumanNet++ \cite{li2025mvhumannet++} & 4500 & \xmark & \xmark & \xmark & \xmark & \xmark & SMPL-X \\
 \rowcolor{green!15} & VolHuMe & 104 & \cmark & \cmark & \cmark & \cmark & \cmark & SMPL-X \\ \hline
\end{tabular}%
}
\caption{A summary of widely-used 3D and 4D human datasets for human reconstruction. VolHuMe is a high-quality dataset featuring real 4D human scans, offering the best trade-off between number of subjects and availability of ground truth data.}
\vspace*{-10pt}
\label{tab:datasets}
\end{table*}

4D human reconstruction aims to generate realistic avatars of humans in motion, capturing geometry, appearance, clothing, and fine-grained articulation such as facial expressions and hand movements. These representations are central to many computer vision and graphics applications, including virtual environments, fashion, and embodied AI.

Data for 4D human reconstruction is commonly collected either in-the-wild using uncalibrated cameras \cite{Patel2021agora,kaufmann2023emdb} or in controlled environments with fixed camera rigs \cite{joo2015panoptic,chatzitofis2020human4d}. In-the-wild datasets often suffer from limited quality due to camera motion, sparse viewpoints, and unconstrained lighting, which has motivated the use of synthetic data to obtain reliable ground truth \cite{black2023bedlam}. Controlled capture setups yield higher-quality data but remain difficult to scale: existing datasets either include many subjects at reduced capture quality or provide high-fidelity geometry and appearance for only a small number of individuals (see Table~\ref{tab:datasets}). Furthermore, most datasets rely on full-body captures from a distance, which limits the reconstruction of fine-grained, closeup details.

Reconstructing high-quality 4D human avatars require multiple complementary data modalities, including RGB images for appearance, depth or point clouds for accurate geometry, and parametric models such as SMPL-X to capture motion and articulation \cite{Pavlakos2019expressive}. However, this combination of detailed geometry, motion, and rich annotations is not consistently available in existing datasets.

To address this gap, we introduce VolHuMe, a large-scale 4D human avatar dataset combining high-resolution textured meshes with extensive ground-truth annotations. VolHuMe features recordings of 104 diverse subjects captured with the Mantis Vision Volumetric Capture System (Fig. \ref{fig:teaser}). The dataset provides a comprehensive range of annotations, including high-resolution meshes and SMPL/SMPL-X fittings, as well as dense point clouds. Unlike existing datasets that typically contain full-body images, VolHuMe captures high-resolution close-ups of all body parts, creating challenging conditions for 3D reconstruction methods. For benchmarking, we evaluate popular approaches using standard metrics, demonstrating VolHuMe’s potential for tasks such as human 3D reconstruction from RGB frames, dynamic mesh sequence reconstruction, and clothed human animation.

 \section{Related Work}
\label{sec:related}

\begin{figure*}[h]
    \centering
    \includegraphics[width=0.9\linewidth]{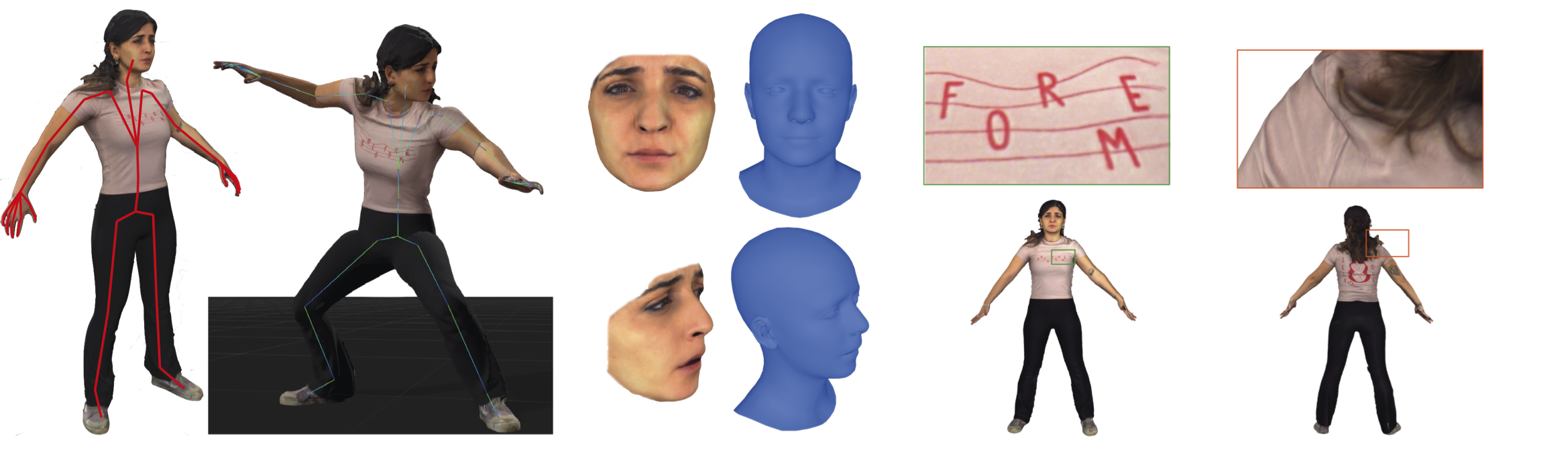}
     \caption{Details of the high-quality data. The rigged mesh (left), alongside with a sample frame from the animation; FLAME model fitting facial features (center); image crops of the high quality mesh showing level of detail and resolution (right).}
    \label{fig:high_quality_bottom}
\end{figure*}

\textbf{4D Human Datasets.}
Deep learning for human pose estimation (HPE), human mesh recovery (HMR), and 4D human modeling depends on datasets that trade off realism, capture conditions, and annotation quality.

In-the-wild datasets (e.g. EMDB \cite{kaufmann2023emdb}) often exhibit noisy imagery and imperfect mesh annotations, while synthetic datasets (e.g., AGORA, BEDLAM) \cite{Patel2021agora,black2023bedlam} provide clean supervision but are limited to parametric representations (SMPL-X) \cite{Pavlakos2019expressive}, which lack fine-grained geometry.

Datasets captured in controlled environments, such as H36M \cite{ionescu2013human3} and Panoptic \cite{joo2015panoptic}, offer improved camera calibration and lighting conditions, enabling high-quality RGB images and meshes. More recent datasets provide high-quality meshes, but lack motion sequences, RGB images, or multi-view data, restricting their applicability to static scenarios.

Several datasets incorporate temporal information. Human4D \cite{chatzitofis2020human4d} provides motion but few subjects, ActorsHQ \cite{icsik2023humanrf} does not provide any parametric body fitting, HUMBI \cite{yoon2021humbi} only provides SMPL fittings, 
AvatarRex \cite{zheng2023avatarrex} and MVHumanNet++ \cite{li2025mvhumannet++} only provides multi-view RGB images and SMPL-X fittings. DNA-Rendering \cite{cheng2023dna} provides cleaned, high-resolution point cloud data but no high-resolution meshes.

Overall, existing resources are often constrained by scale, availability, motion coverage, or geometric detail.
VolHuMe targets these gaps with high-resolution 4D avatars, motion sequences, and comprehensive ground truth.

\textbf{4D Human Reconstruction and Rendering from Multiple Viewpoints.}
Multi-view human reconstruction can be grouped into parametric and non-parametric approaches.
Parametric methods, commonly referred to as HMR, rely on body models such as SMPL and SMPL-X to recover human shape and pose from RGB images.


Non-parametric pipelines reconstruct from calibrated multi-view images via photogrammetry or learned scene representations such as NeRF \cite{tancik2023nerfstudio} and 3D Gaussian Splatting (3DGS) \cite{li2024animatable,zheng2024gps}. Because image-to-geometry inference is ill-posed, these methods commonly leverage additional cues (e.g., masks, parametric initializations, point clouds, or virtual-view supervision) to improve fidelity.

Hybrid approaches combine NeRF or 3DGS representations with parametric models to initialize geometry or pose, balancing computational efficiency and reconstruction detail. Radiant Foam \cite{govindarajan2025radfoam} further couples explicit volumetric geometry with neural fields under a ray-based formulation, enabling realistic camera models and high-quality view synthesis; while not human-specific, it provides a strong high-fidelity baseline in our comparisons.

 \section{Data Acquisition and Processing}
\label{sec:acquisition}

\textbf{Sensing Hardware}
\label{sec:sensing}
The capture area forms a cylindrical volume with a 2.5-meter diameter and a 3-meter height, providing ample space for subjects to move freely. When the participant stands in the area, the distance to the cameras varies from 80 to 120 cm.
The camera rig comprises 32 modules, each equipped with two coplanar RGB cameras and one monochromatic camera for structured light-based depth estimation, totaling 64 RGB cameras (2456 x 2054 pixels) and 32 depth cameras (1280 x 1024 pixels). Unlike other setups \cite{joo2015panoptic,ionescu2013human3,chatzitofis2020human4d,yoon2021humbi,tao2021function4d,zheng2019deephuman,zheng2022structured,icsik2023humanrf,cheng2023dna}, our cameras capture a partial view of the subject; this allows obtaining the highest possible level of detail after merging the views, when compared to full-body shots. 
Cameras are calibrated and hardware-synchronized, recording at 25fps, turning the system into a single capturing device, capable of recording high-fidelity 3D mesh models.

\subsection{VolHuMe Dataset}
\label{sec:volhumedataset}


Each of the 104 diverse captures in VolHuMe consist of a one minute recording, yielding a total of 156,000 annotated frames of 4D human data.

In Fig.~\ref{fig:high_quality_bottom}, we show the high-resolution meshes obtained through volumetric capture, along with animation-ready rigged meshes. Human faces are further fitted to the FLAME model \cite{li2017learning} for enhanced facial accuracy. Accurate SMPL-X fits are also provided as part of the dataset annotations. Following \cite{Patel2021agora}, we assess SMPL-X registration quality using two metrics: the \textit{Skin Error}, computed as the average Euclidean distance between skin vertices of the volumetric scan and the registered SMPL-X surface, and the \textit{Penetrating Clothing Error}, defined as the percentage of clothing vertices from the scan that penetrate the fitted SMPL-X body model. 
Our SMPL-X fits achieve a skin error of 4.73~mm and a penetrating clothing error of 20.05\%. As noted in \cite{Patel2021agora}, an error of approximately 5 mm is significantly below industry standards for measuring live humans; thus, our SMPL-X fits constitute accurate ground truth. For more details please refer to the supplementary material.


\textbf{Data distribution and ethics;} The 104 subjects, all adult participants, are $40 \%$ female and $60 \%$ male, between 18 and 70 years old. Although we did not collect specific ethnicity data to adhere to ethical guidelines, our objective was to ensure high ethnic diversity in the participant pool. All subjects volunteered for the study, signing an information consent form that allowed their data to be used for research purposes and publication. 

 \section{Experiments}

\begin{figure*}[h]
    \centering
    \includegraphics[width=0.98\linewidth]{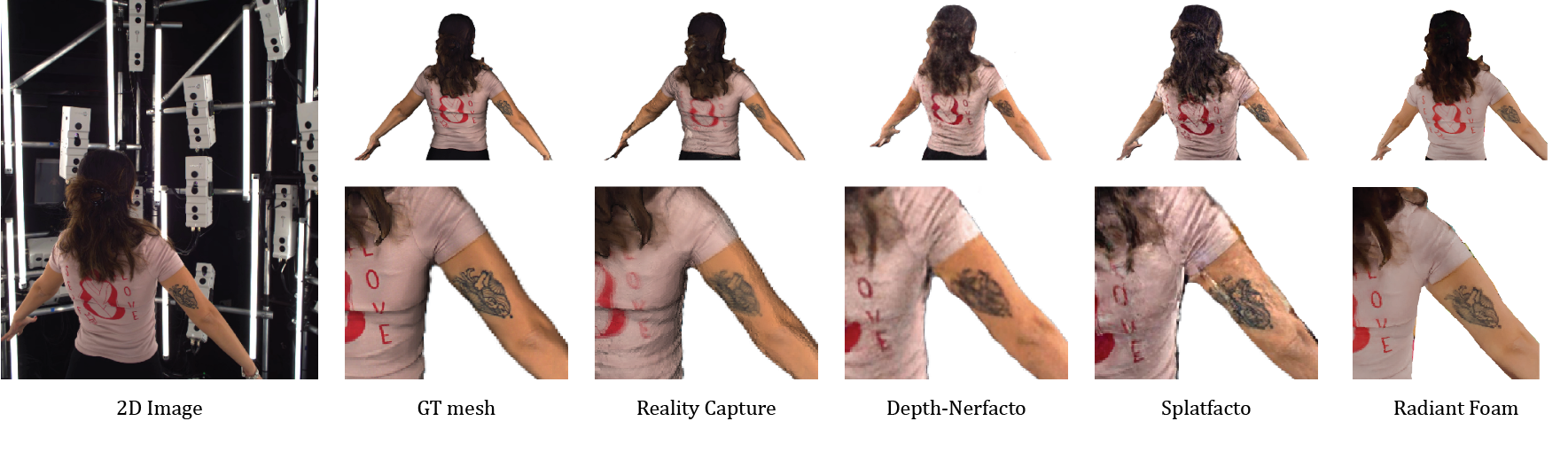}
    \vspace*{-20pt}
    \caption{Comparison of various view synthesis and mesh estimation methods on VolHuMe. While existing methods produce impressive rendering results, they fall short in capturing the finer details compared to our high-resolution ground-truth mesh, highlighting a substantial opportunity for improvement in future research.}
    \label{fig:quality_single_frame}
\end{figure*}

\begin{table}[]
\centering
\resizebox{\linewidth}{!}{%
\begin{tabular}{@{}cc|ccccc@{}}
\toprule
\multirow{2}{*}{\textbf{Method}} &
  \multirow{2}{*}{\textbf{\begin{tabular}[c]{@{}c@{}}NeRF/GS/\\ Photog.\end{tabular}}} &
  \multirow{2}{*}{\textbf{PSNR $\uparrow$}} &
  \multirow{2}{*}{\textbf{SSIM $\uparrow$}} &
  \multirow{2}{*}{\textbf{LPIPS $\downarrow$}} &
  \multicolumn{2}{c}{\textbf{Mesh distance}} \\ \cmidrule(l){6-7} 
                &                                                         &                  &                 &                 & \textbf{Mean}   & \textbf{Std.}   \\ \midrule
Reality Capture & Photog.                                                 & -                & -               & -               & \textbf{2.9357} & \textbf{2.4475} \\
Metashape       & Photog.                                                 & -                & -               & -               & 13.6363         & 19.4009         \\
Meshroom        & Photog.                                                 & -                & -               & -               & \textit{5.4939} & \textit{6.073}  \\
VolNerfacto        & NeRF                                                    & 29.50& 0.967         & 0.041           & 9.3804          & 9.9242          \\
VolNerfacto-big    & NeRF                                                    & 28.48          & 0.856          & 0.057          & 35.6296         & 52.5127         \\
Depth-VolNerfacto  & \begin{tabular}[c]{@{}c@{}}NeRF + \\ depth\end{tabular} & 30.05          & 0.985          & 0.032          & 8.6646          & 7.1147          \\
VolSplatfacto      & 3DGS                                                    & \textit{30.260} & \textit{0.958} & \textit{0.043} & -               & -               \\
VolSplatfacto-big  & 3DGS                                                    & 29.842 & 0.934 & 0.057 & -               & -               \\
Radiant Foam  & -                                                    & \textbf{39.20} & \textbf{0.981} & \textbf{0.016} & -               & -               \\ \bottomrule
\end{tabular}%
}
\caption{View synthesis and mesh estimation results on VolHuMe. In \textbf{bold} the best results, in \textit{italic} the second best. All results are reported in millimeters.}
\vspace*{-10pt}
\label{tab:results_single_frame}
\end{table}

\subsection{View Synthesis and 3D Mesh Estimation}
\label{sec:exp_single}

We evaluate view synthesis and mesh estimation using all views from the first frame of each subject.

Our evaluation includes four categories of methods: photogrammetry-based methods (Reality Capture, Metashape, and Meshroom), NeRF-based methods (Nerfacto, VolNerfacto-big, and Depth-VolNerfacto), 3D Gaussian Splatting-based methods (VolSplatfacto and VolSplatfacto-big), and Radiant Foam \cite{govindarajan2025radfoam}. 
Vanilla NeRF and GS-based methods fail to converge on our data due to the sparsity of viewpoints and the limited visual features in the capture environment, where images share a largely uniform background. To mitigate this issue, we adopt custom variants of these methods by segmenting the human subject and compositing it onto a feature-rich background, which improves view matching and enables stable training, leading to the reported results. A qualitative example of the input data and preprocessing is provided in the supplementary material.

For view synthesis, we use PSNR, LPIPS, and SSIM as evaluation metrics. For mesh estimation, we calculate the mean and standard deviation of point distances to the ground truth mesh. Quantitative and qualitative results are provided in Table \ref{tab:results_single_frame} and Fig. \ref{fig:quality_single_frame}, respectively.

Photogrammetry-based methods achieve the best mesh reconstruction. The results demonstrate improved detail accuracy, which is also visible in the mesh re-projected onto the original image plane. However, the overall detail quality remains lower than that of the ground-truth mesh.

Among the NeRF-based methods, Depth-Nerfacto, that incorporates components from multiple state-of-the-art NeRF approaches and uses depth images, produces the best results. However, these methods tend to over-smooth rendered images, failing to capture the finest details.

3D Gaussian Splatting (3DGS) improves view-synthesis quality over NeRF-based methods, producing sharper appearance and better recovery of mid-frequency textures.

Radiant Foam achieves the best view-synthesis performance in our study, outperforming both NeRF-based methods and 3DGS qualitatively and quantitatively on our metrics. In addition, its renderings exhibit higher perceptual detail than those obtained by photogrammetry when meshes are reprojected to the image plane.

In summary, despite the strong performance of several methods, high-quality reconstruction from sparse multi-view inputs remains challenging. The need for background compositing highlights current limitations of state-of-the-art methods and points to future research directions in 3D human reconstruction.

\begin{figure}[h]
    \centering
    \includegraphics[width=0.75\linewidth]{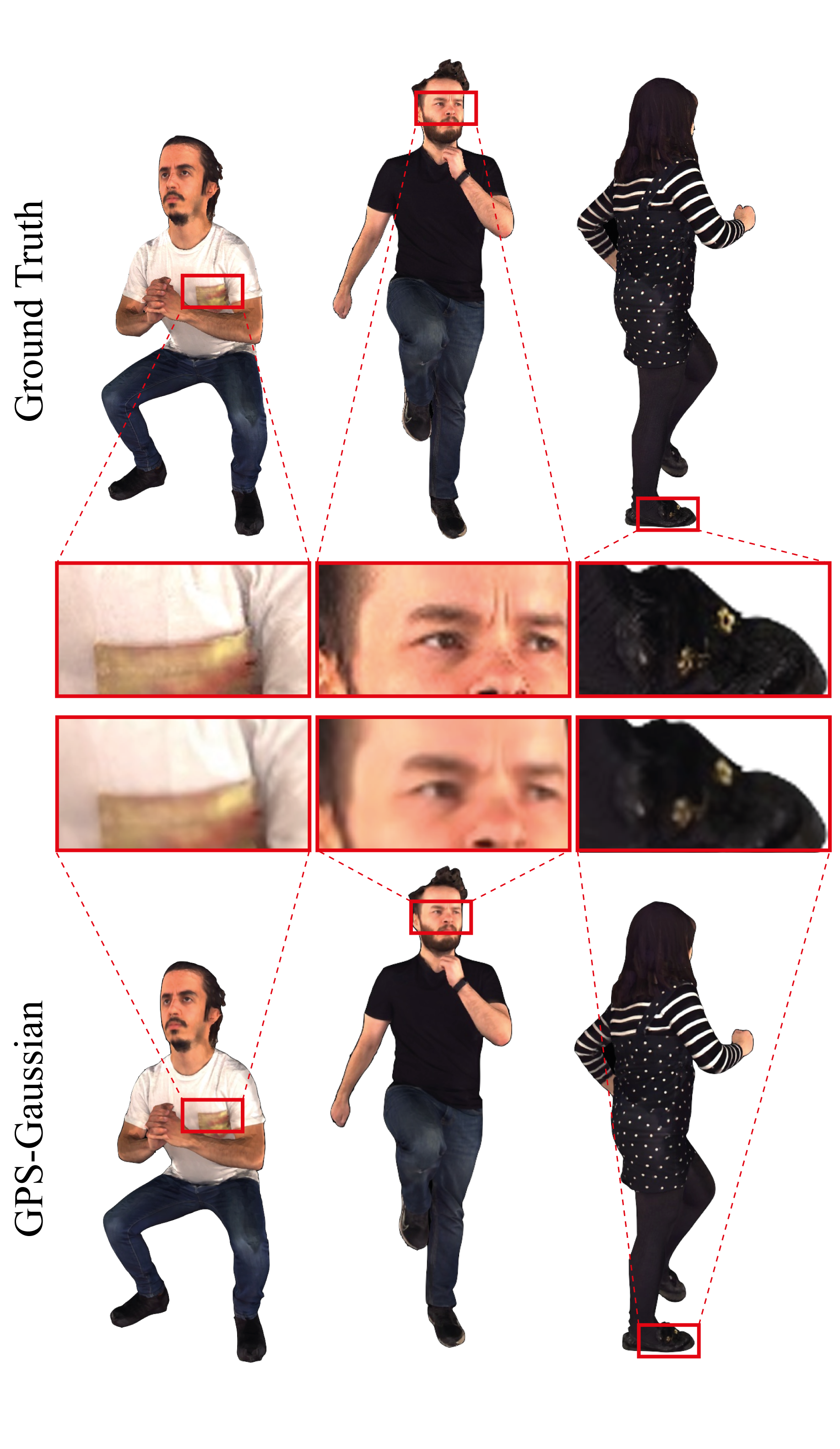}
    \vspace*{-20pt}
    \caption{GPS-Gaussian results on VolHuMe. Here, the mesh resolution had to be reduced, as the method only supports 1024 x 1024 full-body images.}
    \label{fig:quality_gps}
\end{figure}

\begin{figure}[h]
    \centering
    \includegraphics[width=0.85\linewidth]{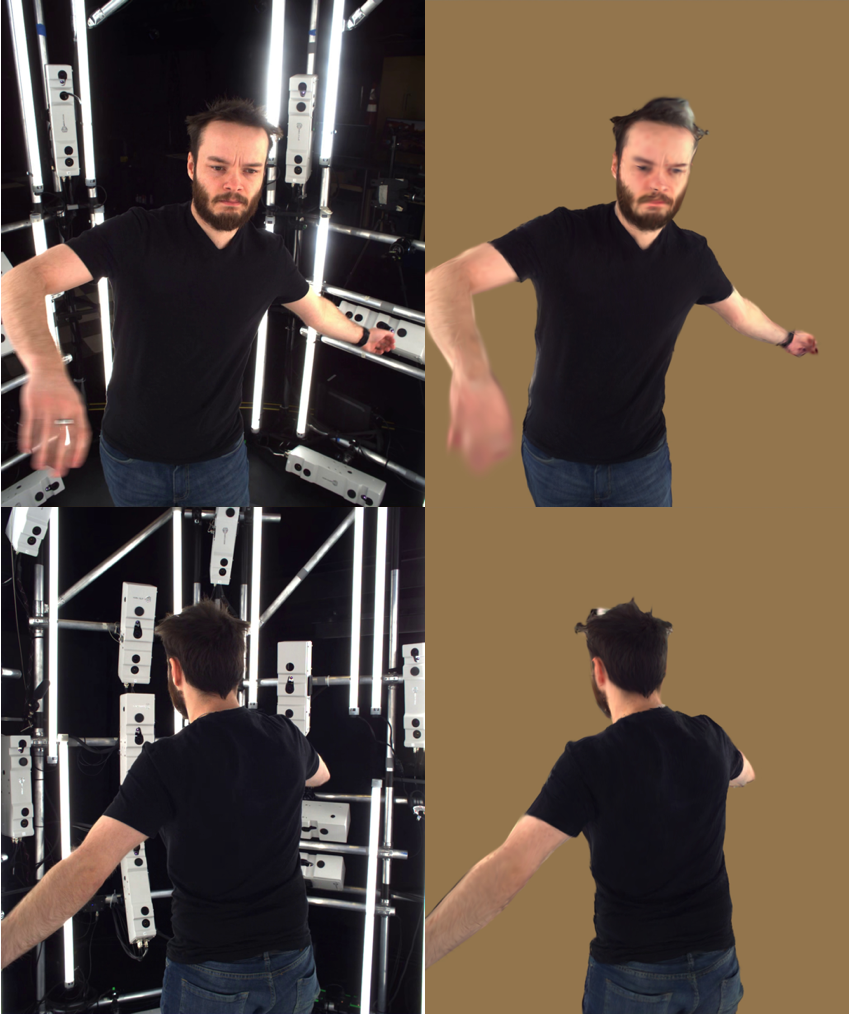}
    \caption{Animatable Gaussians results on VolHuMe: ground truth image (left), reconstruction (right).}
    \label{fig:animatable}
\end{figure}

\subsection{4D Human Reconstruction}

For the 4D human reconstruction task, we evaluate the performance of two state-of-the-art algorithms across the full motion sequence of each subject in VolHuMe. We use GPS-Gaussian \cite{zheng2024gps}, a template-free method that defines Gaussian parameter maps directly on source views, regressing Gaussian properties from those maps, and Animatable Gaussians \cite{li2024animatable}, which leverages the SMPL-X model as a template to parameterize two front-and-back canonical Gaussian maps, where each pixel represents a 3D Gaussian.

We choose PSNR, LPIPS, and SSIM as metrics. Table \ref{tab:results_4d_humans} reports results across many state-of-the-art datasets and VolHuMe, while Figs. \ref{fig:quality_gps} and \ref{fig:animatable} provide qualitative comparisons.

Fig. \ref{fig:quality_gps} illustrates qualitative results of GPS-Gaussians on three actors of the VolHuMe dataset. We follow the standard pipeline proposed in the original work, which reconstructs 4D humans from virtual full-body views rendered at a resolution of 1024$\times$1024 given an input mesh. While this formulation enables stable reconstruction, it does not fully exploit the quality of our high-resolution ground-truth meshes, as distant full-body renderings at limited resolution constrain the recovery of fine geometric and textural details. Quantitatively, results reported in Table \ref{tab:results_4d_humans} show that GPS-Gaussians achieves substantially better performance on VolHuMe than on THuman2.0 across all evaluation metrics. Our gains show that the higher-fidelity meshes in VolHuMe translate into better downstream reconstruction, confirming that VolHuMe provides higher-quality geometric supervision than THuman2.0 even when evaluated with the same reconstruction pipeline.


Animatable Gaussians initializes reconstruction from an SMPL-X model estimated from multi-view full-body RGB images. By normalizing subject height and centering the body in the capture space, this setup simplifies 4D reconstruction and ensures consistent scale across sequences. For our evaluation, we adapted the method to our high-resolution camera setup and to SMPL-X models fitted directly to the reconstructed scan meshes, resulting in variable subject height and increased geometric fidelity. While this adaptation introduces additional complexity, the results reported in Table \ref{tab:results_4d_humans} show that Animatable Gaussians achieves second-best performance on VolHuMe, with results comparable to ActorHQ, despite the latter relying on a dense rig of 160 cameras versus our 32 sparse modules. Notably, our results indicate that sparser camera configurations with close-range views can achieve reconstruction quality on par with dense full-body capture setups. Although our cameras operate at lower resolution, their proximity to the subject enables the capture of finer geometric and textural details. We validate this observation through a controlled experiment in Blender using our reconstructed scan mesh: we compare the effective spatial coverage by counting pixels corresponding to visible reprojected vertices. A 4K full-body image yields 926,774 pixels, whereas our close-up rig achieves 1,582,839 pixels, confirming that partial close-range views provide superior spatial sampling compared to distant full-body shots.


\begin{table}[]
\centering
\resizebox{\columnwidth}{!}{%
\begin{tabular}{cllcccccc}
\hline
\multicolumn{3}{l}{\multirow{2}{*}{Datasets}} & \multicolumn{3}{c}{Animatable Gaussian}              & \multicolumn{3}{c}{GPS-Gaussian}                     \\ \cline{4-9} 
\multicolumn{3}{l}{}                          & PSNR             & SSIM            & LPIPS           & PSNR             & SSIM            & LPIPS           \\ \hline
\multicolumn{3}{c}{ActorHQ}                   & 30.3607          & 0.9682          & 0.0339          & -                & -               & -               \\
\multicolumn{3}{c}{AvatarReX}                 & \textbf{30.6143} & \textbf{0.9803} & \textbf{0.0290} & -                & -               & -               \\
\multicolumn{3}{c}{THuman4.0}                 & 28.0714          & 0.9739          & 0.0515          & -                & -               & -               \\
\multicolumn{3}{c}{THuman2.0}                 & -                & -               & -               & 25.57            & 0.898           & 0.112           \\
\rowcolor{green!15}
\multicolumn{3}{c}{VolHuMe} &
29.5889 & 0.9612 & 0.0316 &
\textbf{36.8513} & \textbf{0.9883} & \textbf{0.0215} \\
\hline
\end{tabular}%
}
\vspace*{-10pt}

\caption{Quantitative results for 4D human reconstruction across multiple state-of-the-art datasets and VolHuMe. We report results for Animatable Gaussians and GPS-Gaussian. In \textbf{bold} the best results.}
\label{tab:results_4d_humans}
\end{table}

 \section{Conclusions}
We presented VolHuMe, a large-scale dataset of high-resolution 4D human meshes acquired with a professional volumetric capture setup and enriched with comprehensive annotations. Unlike existing benchmarks, which rely on full-body capture configurations, VolHuMe adopts a sparse multi-view setup with close-range, partial observations. The experiments show that existing methods still struggle to fully exploit fine-grained geometric and textural detail, particularly under sparse multi-view configurations. Our results indicate that close-range partial observations can achieve higher effective spatial sampling than distant full-body captures, even with fewer cameras. At the same time, the need for auxiliary preprocessing in sparse-view settings highlights current limitations of state-of-the-art reconstruction pipelines. We expect VolHuMe to serve as a challenging benchmark for future work on high-fidelity human reconstruction, rendering and view synthesis.

\section{Acknowledgements}
\label{sec:ack}
This work was carried out with the support of Centre for Immersive Visual Technologies (CIVIT) research infrastructure, Tampere University, Finland\footnote{\url{https://civit.fi/}}.

 \clearpage
\setcounter{page}{1}
\section{Supplementary Material}

\begin{figure*}
    \centering
    \includegraphics[width=0.9\linewidth]{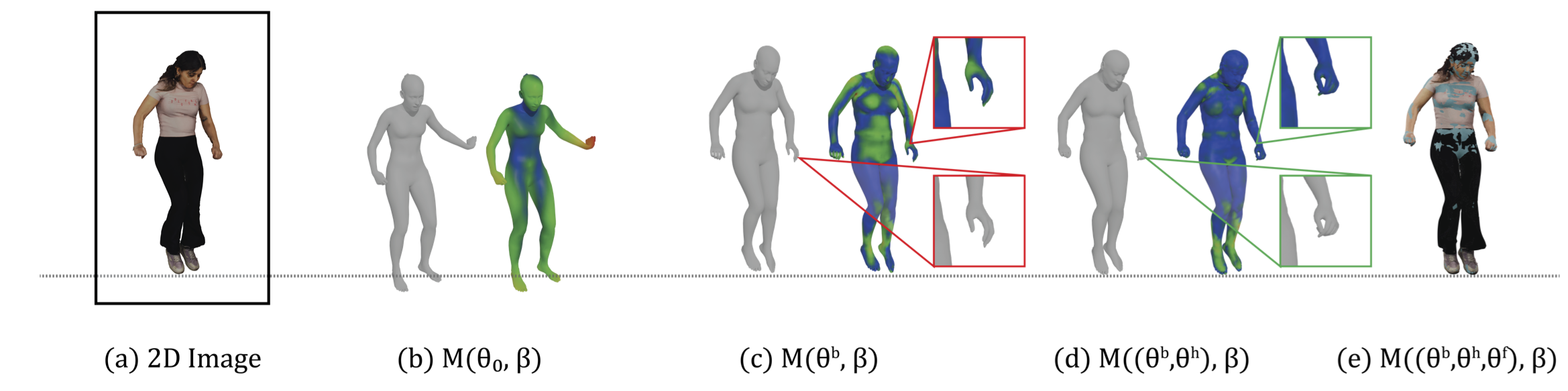}
    \vspace*{-20pt}
    \caption{Steps III-V of our SMPL-X semi-automatic registration process. In each pair of images, the gray mesh represents the SMPL-X model; the colored mesh visualizes the distance between points on the SMPL-X and the original high-resolution mesh; deeper blue indicates higher fitting accuracy. In Step III, starting from a 2D image (a), an estimator from 2D images \cite{choutas2020monocular} provides an initial set of parameters, $M(\theta_0, \beta)$, which appears misaligned (b). In Step IV, the optimizer adjusts the body pose and shape parameters to refine the fit to $M(\theta^b, \beta)$, excluding hands and facial details (c). During Step V, ICP is used to produce $M((\theta^b, \theta^h), \beta)$, an SMPL-X mesh with hands and shape further refined but without final parameters (d). Finally, inverse LBS yields the parameters of the light blue mesh $M((\theta^b, \theta^h, \theta^f), \beta)$, which aligns closely with the original mesh (e).}
    \label{fig:fitting}
\end{figure*}

\subsection{Semi-automatic Human Parametric Model Registration}
\label{sec:fitting}

As discussed in Sec. 2, most existing methods for 4D human registration rely on parametric mesh initialization using models such as SMPL or SMPL-X. In our dataset, we introduce a novel, highly accurate method for the semi-automatic registration of SMPL-X to align the high-resolution, high-fidelity mesh $\mathcal{M}$ for each frame.

A SMPL-X model is represented by a function $M(\theta, \beta, \psi)$, where $\theta$ controls the pose, $\beta$ defines the shape, and $\psi$ represents facial expressions. The pose parameters $\theta = (\theta^b, \theta^f, \theta^h)$ are further broken down into body pose $\theta^b$, facial pose $\theta^f$, and hands pose $\theta^h$.

The body fitting procedure we propose is detailed in Algorithm \ref{alg:registration} and Steps III to V are illustrated in Fig. \ref{fig:fitting}. Each processing step is detailed below.

\begin{algorithm}
  \caption{The goal is to obtain a SMPL-X mesh $M((\theta^b,\theta^h,\theta^f), \beta,\psi)$ that fits the high-resolution mesh $\mathcal{M}$. }
  \begin{enumerate}[label=\Roman*.]
  
  \item $ICP+LBS(\mathcal{M}_a,M(\theta_t,\beta_t), \mathcal{P}_a) \rightarrow M(\theta_a, \beta_a)$ : intial SMPL-X estimate in A-pose
  
  
  \item $E(\beta \leftarrow \beta_a) \rightarrow \beta$ : shape estimation of naked body parameters
  
  \item $\forall (2D image, \beta) \rightarrow M(\theta_0, \beta)$ : per-frame SMPL-X body pose estimate
  
  \item $\mathcal{L}(\mathcal{M}, M(\theta_0, \beta)) \rightarrow M(\theta^b, \beta)$ : pose optimization without hands and facial features
  
  
  \item $ICP+LBS(\mathcal{M},M(\theta^b, \beta),\mathcal{P}_h) \rightarrow M((\theta^b,\theta^h), \beta)$ : final accurate SMPL-X mesh without face features
  
  \item $(M((\theta^b,\theta^h), FLAME (\mathcal{M})) \rightarrow M((\theta^b,\theta^h,\theta^f), \beta,\psi)$ : final accurate SMPL-X mesh
  
\end{enumerate}
  \label{alg:registration}
\end{algorithm}

\textbf{Step I. Initial body shape estimation.} 

As $\mathcal{M_A}$ denotes a clothed individual captured in A-pose, we estimate the body shape underneath clothing. We manually select a set of 15 points $\mathcal{P} : {p_i = (x_i, y_i, z_i)}$ on the surface of the scan mesh. The points in $\mathcal{P}_a$ correspond to prominent anatomical landmarks located at key joints on principal limbs and torso, consistently identified on both the high-resolution mesh $\mathcal{M_A}$ and the SMPL-X model in A-pose. 

Next, we apply an Iterative Closest Point (ICP) algorithm between the template SMPL-X mesh $M(\theta_t, \beta_t)$ and the high-resolution mesh $\mathcal{M_A}$, using the points $\mathcal{P}_a$ as constraints. This process results in a well-aligned SMPL-X mesh $M(\theta_a, \beta_a)$. However, since ICP only aligns the mesh vertices, the underlying SMPL-X parameters $(\theta_a, \beta_a)$ remain unknown. To recover them, we apply inverse Linear Blend Skinning (LBS), aligning a SMPL-X template to our mesh. The resulting $M(\theta_a, \beta_a)$ with known parameters is fitted to the clothed human, rather than to the body beneath the clothes.

\textbf{Step II. Precise body shape.} To obtain the parameters for the naked body we employ an energy minimization approach, as in \cite{estimating_body_shape}:
\begin{equation}
E(\beta) = E_{fit}(\beta) + \lambda_{\beta}E_\beta(\beta)
\end{equation}
where $E_{fit}$ is a fitting term, and $\lambda_\beta E_\beta$ serves as a regularization term for the shape. 
The fitting process begins with $\beta = \beta_a$ and results in a refined estimate of the shape parameters $\beta$ in A-pose. Given the accurate $\beta$, in the subsequent stages our goal is to retrieve the body pose $\theta^b$, hand $\theta^h$ and facial $(\theta^f,\psi)$ parameters . 


\textbf{Step III. Body pose estimation for each frame.} We render a single view of the high-resolution textured mesh for each frame. The image is fed into a standard SMPL-X estimator \cite{choutas2020monocular}. This estimator provides an initial guess of the pose parameters $\theta_0$ and we obtain $M(\theta_0, \beta)$, where $\beta$ is the output of the previous step.

\textbf{Step IV. Body pose optimization.} We employ an Adam optimizer to fit the parametric mesh $M(\theta_0, \beta)$ to the high-resolution mesh $\mathcal{M}$. This is achieved by optimizing the body pose parameters $\theta^b$.
The optimizer is trained for 500 steps with a learning rate of $1e-2$ to minimize the following loss:

\begin{equation}
    \mathcal{L}(\mathcal{M},M(\theta_0, \beta)) = \mathcal{L}_{CD} + \mathcal{L}_N + \mathcal{L}_{MSE}
\end{equation}

where $L_{CD}$ represents the Chamfer Distance (CD) between the vertices of the parametrized and high-res meshes, $L_N$ is a normal consistency loss between the faces of the two meshes, and $L_{MSE}$ is the Mean Square Error (MSE) between the foot joints of $M(\theta_0, \beta)$ and the estimate $M(\theta^b, \beta)$. While $\mathcal{L}_{CD}$ and $\mathcal{L}_N$ focus on fitting the body pose and shape, $\mathcal{L}_{MSE}$ ensures accurate alignment of the feet to the ground plane. 

The output is a refined SMPL-X mesh $M(\theta^b, \beta)$, where the overall body pose and shape are accurately fitted. However, the facial expression and hand pose remain as initially estimated by the 2D detector, as they have not yet undergone any further refinement.

\textbf{Step V. Hands refinement.} Accurately fitting hand poses is particularly challenging due to the small size of the hands, their complex articulation, and frequent self-occlusions. Even state-of-the-art volumetric capture systems struggle to precisely capture hand motion due to resolution limitations and the lack of sufficient visual cues. 

When the estimated hand position of the SMPL-X mesh is close to the corresponding hands in $\mathcal{M}$, also in this case we apply an Iterative Closest Point (ICP) algorithm between the refined SMPL-X mesh $M(\theta^b, \beta)$ and the scan $\mathcal{M}$, without using control points. However, if the estimated hand pose deviates significantly from the ground-truth scan, we manually select 10 key points $\mathcal{P}_h$ on the fingertips from both the scan and the refined SMPL-X mesh. We then apply ICP using $\mathcal{P}_h$ as guiding references, yielding a well-aligned SMPL-X mesh $M((\theta^b,\theta^h), \beta)$ with unknown parameters $(\theta^b,\theta^h)$. As in Step I, to recover such parameter, we apply inverse LBS between $M((\theta^b,\theta^h), \beta)$ and $M(\theta^b, \beta)$, which is the closest SMPL-X mesh for which we already know the parameters. This results in the final SMPL-X mesh $M((\theta^b,\theta^h), \beta)$, with accurately posed and detailed hand features.

\textbf{Step VI. Face Fitting.} Face fitting is performed using the FLAME model. For fitting, we adhered to the official FLAME pipeline: we first extracted vertices corresponding to the facial region from the scan, and then employed 51 predefined 3D facial landmarks to initialize the fitting process, ensuring accurate alignment between the model and the captured facial data. The output of the face fitting process includes the facial pose parameters $\theta^f$ and expression parameters $\psi$ , which are subsequently integrated with the previously fitted SMPL-X model. Overall, these steps yield a fully accurate SMPL-X mesh $M((\theta^b,\theta^h,\theta^f),\beta,\psi)$.

\subsection{SMPL-X Fitting Accuracy}
\begin{figure}
\centering
\includegraphics[width=0.8\linewidth]{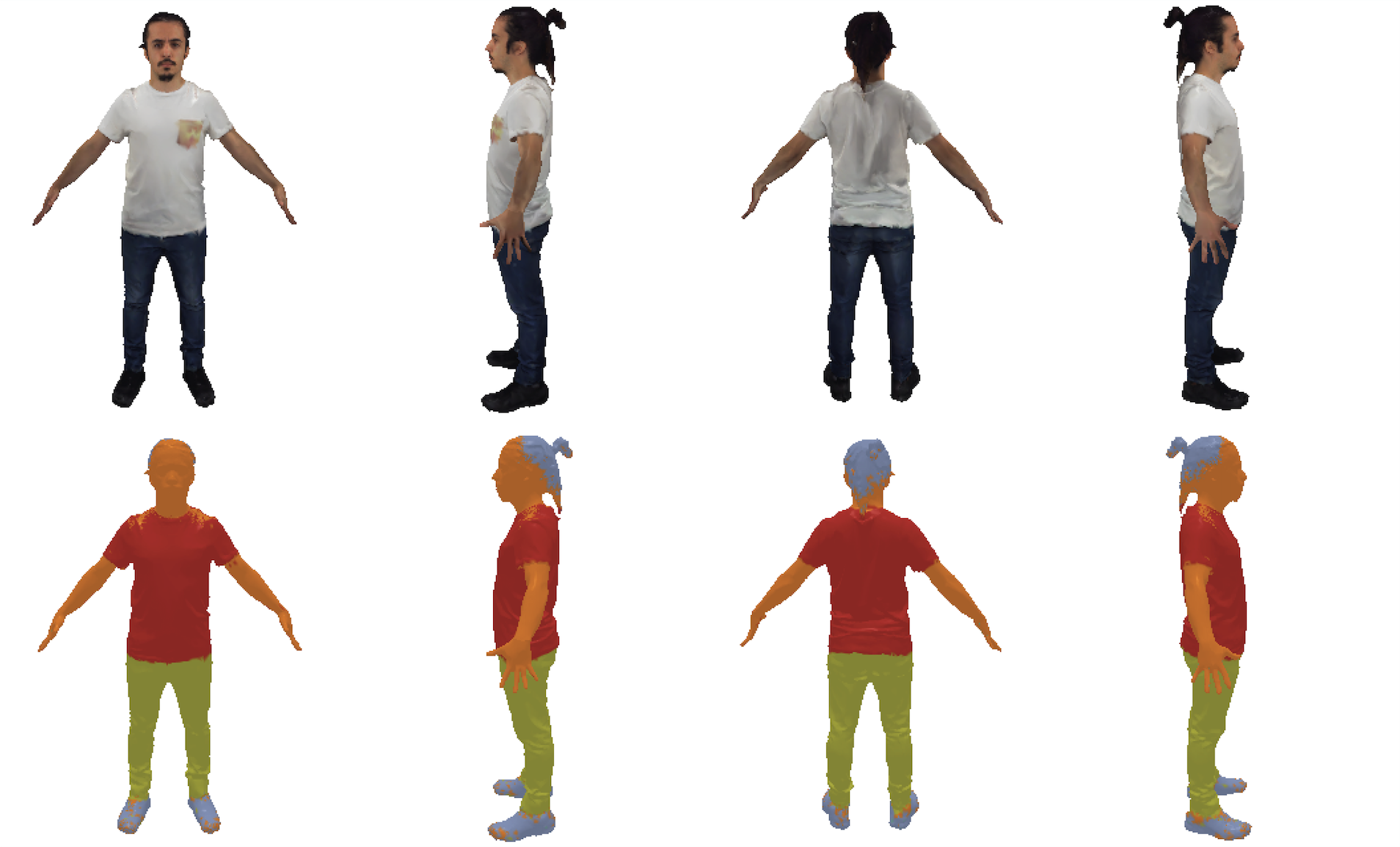}
\caption{Example of segmentation on our scans. The points corresponding to the visible skin can be used to compute a metric for the accuracy of our registration pipeline.}
\label{fig:garments}
\end{figure}

As discussed in Sec. 3.1, we evaluate the registration quality of the SMPL-X model against the scan using two metrics: Skin Error and Penetrating Clothing Error. Computing these metrics requires a 3D semantic segmentation of the reconstructed meshes. To this end, we apply the method of \cite{antic2024close} to our high-resolution meshes to obtain an accurate ground truth segmentation into skin, hair, and clothing regions. The segmented skin vertices are then used to compute the registration accuracy metrics reported in Table \ref{tab:fitting_accuracy}. Qualitative examples of the resulting garment segmentation are shown in Fig.~\ref{fig:garments}.

\begin{figure}
    \centering
    \includegraphics[width=.8\linewidth]{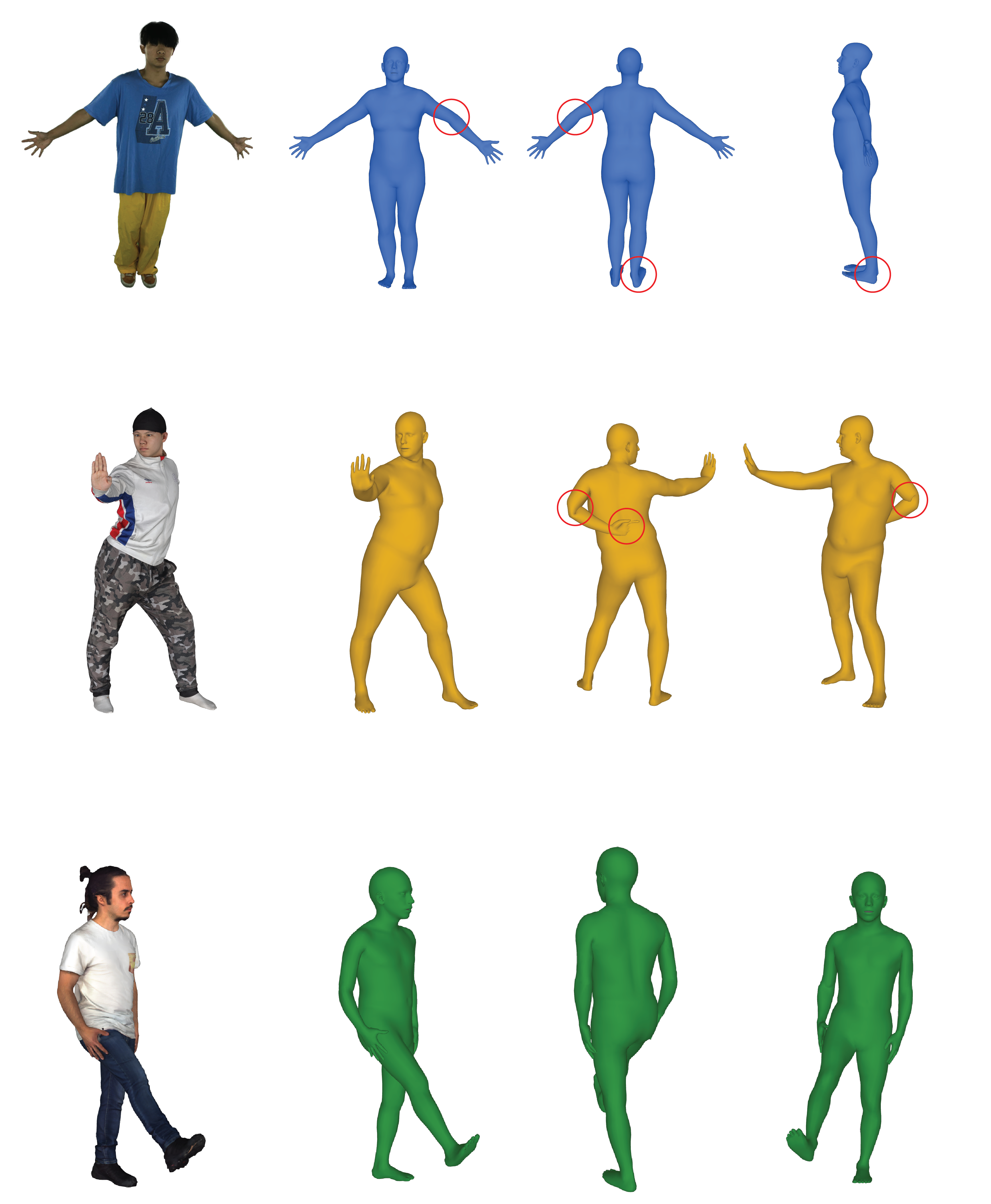}
    \caption{\textbf{Top to bottom}: MVHumanNet, THuman and VolHuMe datasets. \textbf{Left to right}: Ground truth textured mesh (sample image for MVHumanNet, which has no textured meshes), same view SMPL-X fitting, back view, side view. We highlight in red some issues of other datasets, such as unnatural bends and mesh inter-penetrations. Our SMPL-X fitting pipeline solves most of the issues present in the other datasets.}
    \label{fig:comparison_datasets_2}
\end{figure}

 In Table \ref{tab:fitting_accuracy}, we present the accuracy results for each step of our registration pipeline. For the shape estimation $\beta$, the best result is achieved in Step II, where we obtain the lowest penetrating clothing error (13.4\%) with a skin error of 4.52 mm. For the body pose estimation $\theta$, the lowest clothing error (19.06\%) occurs at Step IV, where only the body pose has been optimized, resulting in a relatively higher skin error of 7.47 mm, as neither the face nor the hands were fitted. However, when both face and hands are also optimized (Step VI), the penetrating clothing error remains very close (20.05\%) but with a significantly improved skin error of 4.73 mm, on par with the state-of-the art AGORA dataset. As noted in the AGORA paper, an error of approximately 5 mm is significantly below industry standards for measuring live humans; thus, our SMPL-X fits constitute valid pseudo ground truth. A qualitative comparison against other state-of-the-art datasets providing SMPL-X registrations is shown in Fig.~\ref{fig:comparison_datasets_2}, where our fits exhibit noticeably fewer artifacts—such as unnatural joint bending and mesh interpenetrations—than competing methods.
\textbf{Ablation study.} In \textbf{Ablation A}, we performed the registration without optimizing the shape parameters ($\beta$) under clothing. Although the skin error decreased to 5.31 mm by the final step, the penetrating clothing error remained high (44.68\%), highlighting the critical role of Step II in obtaining a properly sized model fully contained within the scan mesh. \textbf{Ablation B} skipped Step IV, retaining the initial body pose estimated by a 2D detector. This resulted in a significant skin error (160.45 mm), indicating that without proper body pose initialization, subsequent optimization (particularly for hands) might incur in a local minimum. 
\textbf{Ablation C}, which omitted both Steps II and IV, showed similarly poor results with an even higher penetrating clothing error, emphasizing the importance of each pipeline step and demonstrating how parameter initialization critically affects optimization quality.

\begin{table}[]
\centering
\resizebox{0.7\columnwidth}{!}{%
\begin{tabular}{cccc}
\hline
\textbf{Pipeline}                                                                             & \textbf{Step} & \textbf{\begin{tabular}[c]{@{}c@{}}Skin\\ (mm)\end{tabular}} & \textbf{\begin{tabular}[c]{@{}c@{}}Over clothes\\ (\%)\end{tabular}} \\ \hline
\multirow{3}{*}{\textbf{Shape}}                                                               & I             & \textbf{4.31}                                                & 52.24                                                                \\
                                                                                              & II            & 4.52                                                         & \textbf{13.4}                                                        \\
                                                                                              & III           & 62.69                                                        & 44.05                                                                \\ \hline
\multirow{3}{*}{\textbf{Body}}                                                                & IV            & 7.47                                                         & \textbf{19.06}                                                       \\
                                                                                              & V             & 4.85                                                         & 23.42                                                                \\
                                                                                              & VI            & \textbf{4.73}                                                & 20.05                                                                \\ \hline
\multirow{2}{*}{\textbf{\begin{tabular}[c]{@{}c@{}}Ablation A\\ (skip step II)\end{tabular}}} & IV            & 10.33                                                        & 40.95                                                                \\
                                                                                              & V             & 5.31                                                         & 44.68                                                                \\ \hline
\textbf{\begin{tabular}[c]{@{}c@{}}Ablation B\\ (skip step IV)\end{tabular}}                  & V             & 160.45                                                       & 66.76                                                                \\ \hline
\textbf{\begin{tabular}[c]{@{}c@{}}Ablation C\\ (skip steps II, IV)\end{tabular}}             & V             & 147.54                                                       & 70.87                                                                \\ \hline
\end{tabular}
}
\caption{SMPL-X registration accuracy results and ablations.}
\vspace*{-10pt}
\label{tab:fitting_accuracy}
\end{table}

\subsection{Custom view synthesis methods}

\begin{figure}
\centering
\includegraphics[width=0.8\linewidth]{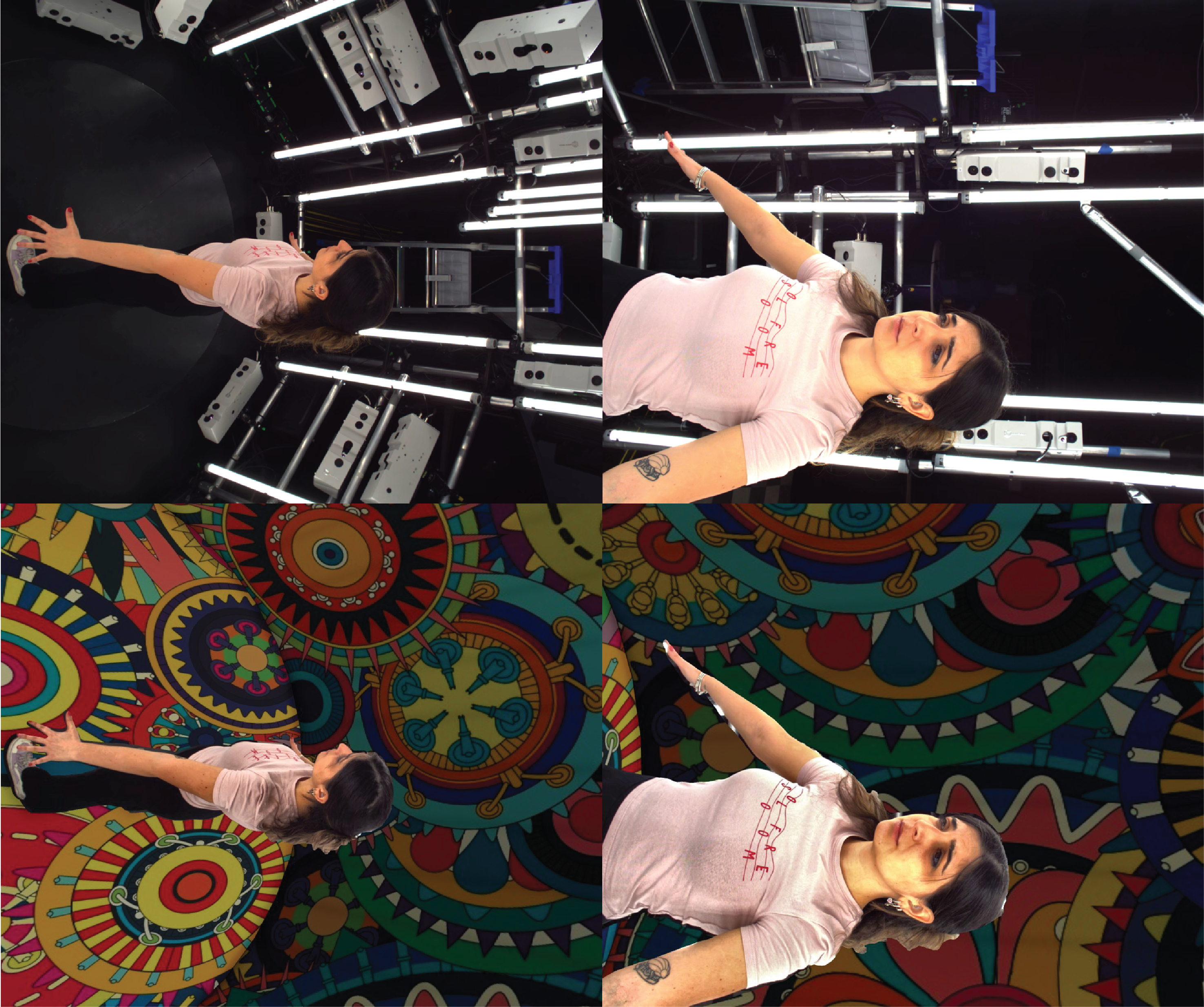}
\caption{Background replacement for NeRF- and GS-based methods.}
\label{fig:background_substitution}
\end{figure}

As discussed in Sec. 4.1, vanilla NeRF- and GS-based methods fail to converge on VolHuMe due to the sparse camera configuration and the limited visual features of the capture environment. The volumetric capture setup provides accurate segmentation of the subject within the acquisition space. We leverage this information, together with the calibrated camera parameters, to segment the human subject in each view and replace the original background with a feature-rich one. This preprocessing improves view matching across sparse viewpoints and enables stable training of NeRF- and GS-based models. Qualitative examples of the input data and the background replacement process are shown in Fig.~\ref{fig:background_substitution}.

\begin{figure*}
    \centering
    \includegraphics[width=.73\linewidth]{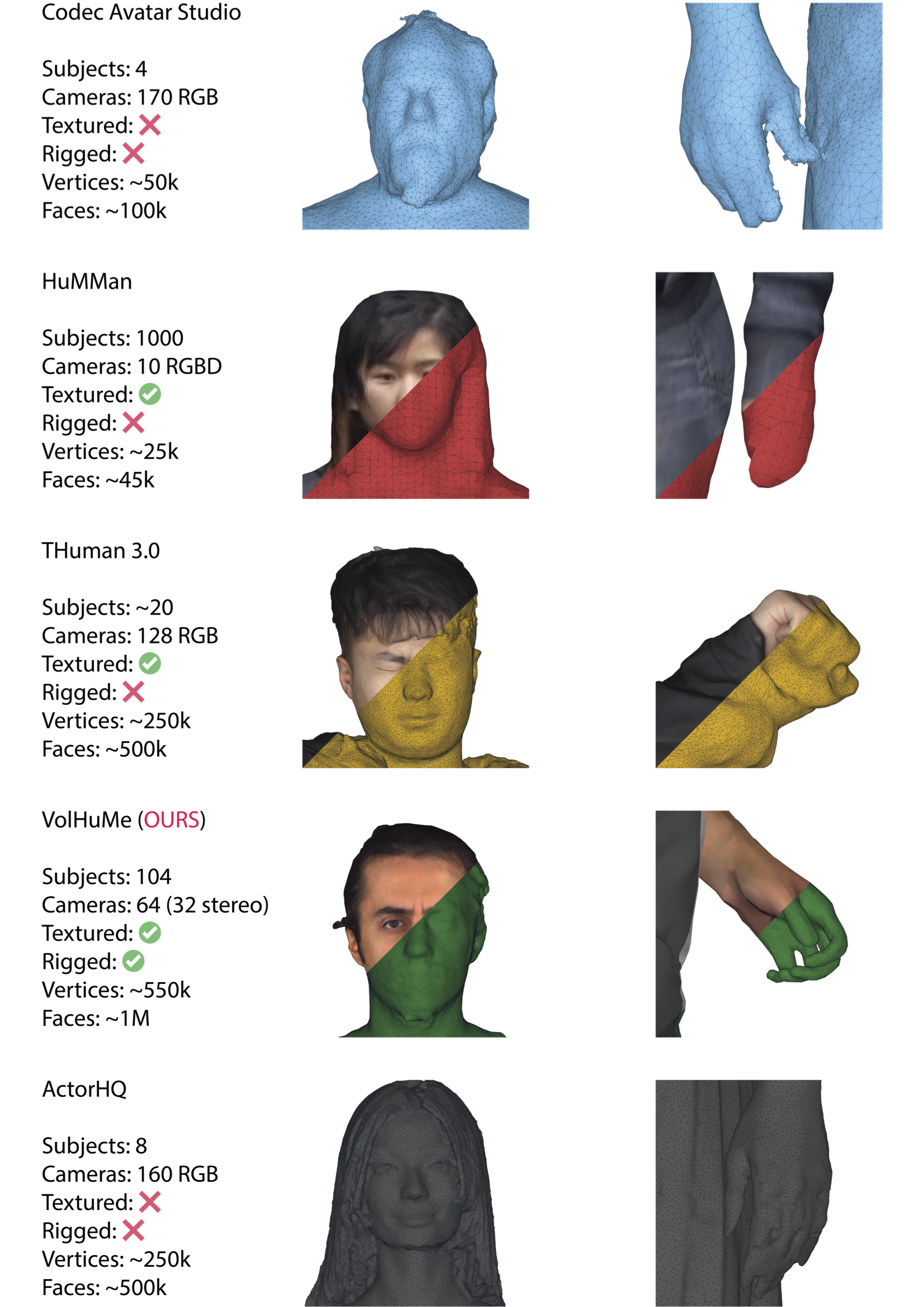}
    \caption{Comparison with other state-of-the-art datasets. With VolHuMe, we offer a good trade-off between number of subjects, size and overall quality of the mesh, that is "more Delaunay" with respect to the other datasets, and the texture resolution. Moreover, our dataset is the only one to offer both textured and rigged characters, ready for animation.}
    \label{fig:comparison_datasets}
\end{figure*}

\newpage

\bibliographystyle{IEEEbib}
\bibliography{refs}

\end{document}